\documentclass[useAMS,usenatbib,usegraphicx]{mn2e}

\usepackage{rotating}
\usepackage{lscape}
\usepackage{color}

\newcommand{\be}{\begin{equation}}
\newcommand{\ee}{\end{equation}}
\newcommand{\bea}{\begin{eqnarray}}
\newcommand{\eea}{\end{eqnarray}}

\def\ifm#1{\relax\ifmmode#1\else$\mathsurround=0pt #1$\fi}
\def\kms{\ifmmode\,{\rm km}\,{\rm s}^{-1}\else km$\,$s$^{-1}$\fi}

\def\r200{r_{200}}
\def\m200{m_{200}}

\title[Orbits of Simulated Satellites]{The Orbital Ellipticity of
  Satellite Galaxies and the Mass of the Milky Way} \author[C. Barber
et al.]  {Christopher Barber$^{1}$, Else Starkenburg$^{1,2}$, Julio F.
  Navarro$^{1,3}$, Alan W. McConnachie$^{4}$
  \newauthor 
and Azadeh Fattahi$^{1}$\\
$^{1}$Dept. of Physics and Astronomy, University of Victoria, P.O. Box 1700, STN CSC, Victoria BC V8W 3P6, Canada \\
$^{2}$CIFAR Global Scholar and CITA National Fellow\\
$^{3}$CIFAR Senior Fellow \\
$^{4}$NRC Herzberg Institute of Astrophysics, 5071 West Saanich Road, Victoria,
British Columbia, Canada, V9E 2E7}

\begin{document}


\pagerange{\pageref{firstpage}--\pageref{lastpage}} \pubyear{2013}

\maketitle

\label{firstpage}

\begin{abstract} We use simulations of Milky Way-sized dark matter
  haloes from the Aquarius Project to investigate the orbits of
  substructure haloes likely, according to a semi-analytic galaxy
  formation model, to host luminous satellites. These tend to populate
  the most massive subhaloes and are on more radial orbits than the
  majority of subhaloes found within the halo virial radius. One 
  reason for this (mild) kinematic bias is that many low-mass
  subhaloes have apocentres that exceed the virial radius of the main
  host; they are thus excluded from subhalo samples identified within
  the virial boundary, reducing the number of subhaloes on radial
  orbits. Two other factors contributing to the difference in orbital
  shape between dark and luminous subhaloes are their dynamical
  evolution after infall, which affects more markedly low-mass (dark)
  subhaloes, and a weak dependence of ellipticity on the redshift of
  first infall. The ellipticity distribution of luminous satellites
  exhibits little halo-to-halo scatter and it may therefore be
  compared fruitfully with that of Milky Way satellites. Since the
  latter depends sensitively on the total mass of the Milky Way we can
  use the predicted distribution of satellite ellipticities to place
  constraints on this important parameter. Using the latest estimates
  of position and velocity of dwarfs compiled from the literature, we
  find that the most likely Milky Way mass lies in the range
  ${6\times 10^{11} \, M_\odot <M_{200}<3.1 \times 10^{12} \,
    M_\odot}$, with a best-fit value of ${M_{200}=1.1\times 10^{12}
    M_\odot}$. This value is consistent with Milky Way mass estimates
  based on dynamical tracers or the timing argument. \end{abstract}


\section{Introduction}
\label{SecIntro}

Satellite galaxies have long been used as kinematic tracers of the
gravitational potential of the Milky Way (MW) halo
\citep[e.g.,][]{Hartwick1978,LyndenBell1983,Zaritsky1989,Kulessa1992,
  Kochanek1996,Wilkinson1999,Battaglia2005,Sales2007a,Boylan2013}.
The usefulness of this technique, however, has been traditionally
limited by the relatively small number of satellites known, by
uncertainties in their estimated distances, and by the availability of
a single component of the orbital velocity, along the line of
sight. This state of affairs, however, is starting to change.

Over the last decade, surveys like the Sloan Digital Sky Survey (SDSS)
have mapped { large areas of the sky}, an effort that has led to the
discovery of a number of very faint satellite galaxies (the
``ultra-faint'' dwarf spheroidal companions of the Milky Way) whose
star formation history, chemical evolution, mass, distance, and
velocity have now been estimated through deep follow-up observations
\citep[e.g.,][]{Willman2005,Zucker2006a,Zucker2006b,Belokurov2007,Walsh2007,Irwin2007,Kirby2008,Martin2008,Aden2009,Norris2010,Wolf2010,Simon2011,Brown2012}. Distance
estimates have also improved, to the point that the distances to most
satellites are now known to better than $\sim 10\%$ from measurements
of resolved stellar populations. Further, the superior angular
resolution of the Hubble Space Telescope has enabled proper motion
estimates for nearby dwarfs from images with a time baseline of just a
few years \citep[e.g.,][]{Piatek2002}, and modern adaptive optics
systems promise to reach comparable angular resolution from the ground
\citep[e.g.,][]{Rigaut2012}. Finally, in the near future, a great leap
forward is expected from the Global Astrometry Interferometer for
Astrophysics (Gaia) satellite \citep[e.g.,][]{Lindegren1996}. This
mission is expected to measure the proper motions of the MW dwarf
spheroidal system to an accuracy of a few to tens of \kms, depending
on the satellite's properties \citep{Wilkinson1999}.

Accurate proper motions, radial velocities, positions, and distances
can be turned into satellite orbits after assuming a mass profile for
the Galaxy. The { shapes} of these orbits are expected to contain
information about the circumstances of the accretion of individual
satellites, as well as about the evolution of the potential well of
the Galaxy over time.  Decoding such information, however, is not
straightforward, and is best attempted by contrasting observations with
realistic simulations that resolve in detail the dynamical evolution
of the potential sites of dwarf galaxy formation. 

Although there are in the literature a number of studies of the
kinematics of satellite systems and their relation to the haloes they
inhabit \citep[e.g.,][]{Tormen1997, Tormen1998, Ghigna1998, vandenBosch1999, Balogh2000, Taffoni2003, Kravtsov2004, Gill2004, Gill2005,Diemand2007, Sales2007a, Ludlow2009}, most have
dealt primarily with the orbits of substructure haloes (referred to
hereafter as subhaloes) in general. Luminous satellites inhabit a small
fraction of subhaloes, and their orbits might therefore very well be
substantially biased relative to those of typical subhaloes. Making
progress demands not only simulations with numerical
resolution high enough to resolve all potential sites of luminous
satellite formation but also a convincing way of pinpointing the few
subhaloes where those satellites actually form.

A number of simulations that satisfy the numerical resolution
requirement have been recently completed, notably the { six Milky
Way-sized haloes of the Aquarius Project}
\citep[]{Springel2008}, as well as the Via Lactea II halo
\citep{Diemand2008}, and its higher-resolution version GHALO
\citep{Stadel2009}. In this study we combine the Aquarius Project
haloes with the semi-analytical model of \citet{Starkenburg2013} to
identify satellites with luminosities down to the ``ultra-faint''
regime. We study the orbital
distribution of these satellites, and explore its dependence on
satellite properties such as stellar mass and accretion time. Our
analysis yields predictions that should prove useful in the near
future, when Gaia delivers accurate 6D phase space information for many
Milky Way satellites. We describe here a possible application, making
use of published proper motions, positions and radial velocities
of the most luminous Milky Way satellites to constrain the mass of the
Milky Way halo.

The plan of this paper is as follows. In Sec.~\ref{SecSimSats} we
describe the simulated satellite sample we use, together with a brief
discussion of the numerical simulations and of the semi-analytic
galaxy formation model adopted. We describe the analysis techniques
used to compute orbital properties for satellites and subhaloes and
present their orbital ellipticity distributions in
Sec.~\ref{SecResults}. We investigate in the same section the origin
of their differences before, finally, in Sec.~\ref{SecCompMW},
comparing the orbits of simulated dwarf galaxies to those of MW dwarfs
in order to discuss the constraints they imply on the total virial mass of
the Milky Way.  We summarize our main conclusions in
Sec.~\ref{SecConc}.

\section{Simulated Satellites}
\label{SecSimSats}

\subsection{The N-body Simulations}

The Aquarius Project consists of a suite of six high-resolution dark
matter-only simulations of haloes with virial mass\footnote{We define
  all halo ``virial'' quantities (labelled with a ``200'' subscript)
  as those measured within a sphere of mean density 200 times the
  critical density for closure, $\rho_{\rm crit}$.} $M_{200}$ { in
  the range ${(0.8 - 1.8) \times 10^{12} M_\odot}$}. We use the level-2
resolution runs of { all six Aquarius haloes} (named Aq-A through
Aq-F), resolved with several hundred million particles each
\citep[see][for details]{Springel2008}. 
The numerical resolution of the level-2 Aquarius haloes allows us to
track dark matter haloes with masses as small as $10^5 \, M_\odot$,
which we identify and track using the groupfinder {\sc subfind}
\citep{Springel2005}. This algorithm recursively identifies all
self-bound substructures with at least 20 particles present within a
set of haloes first identified using a standard friends-of-friends
technique \citep{Davis1985}.

All simulations assume a ``standard'' $\Lambda$CDM cosmogony, with the
same parameters as the Millennium Simulation \citep{Springel2005}:
$\Omega_{\rm M}=0.25$, $\Omega_{\Lambda}=0.75$, $n=1$, $h=0.73$, and
$\sigma_8=0.9$. { Although these parameters are now out of favour
considering the recently published results from the Planck satellite
\citep{Planck2013}, we expect them to have little effect on the
detailed non-linear structure and substructure of dark matter haloes,
which concern us here \citep[see,
e.g.,][]{Wang2008,BoylanKolchin2010,Guo2013}}.

\subsection{The semi-analytic model}
\label{SecSAM}

A semi-analytic model of galaxy formation is grafted on to the evolving
collection of {\sc subfind} haloes and subhaloes linked as a function of
time by a merger tree \citep{Baugh2006, Benson2010}. The particular
model implementation we use here is described by
\citet{Starkenburg2013} and is an extension of earlier work
\citep{Kauffmann1999, Springel2001, DeLucia2004, Croton2006,
  DeLucia2007, DeLucia2008, Li2009, Li2010}. The main ingredients for
the model are analytic prescriptions for gas-cooling, re-ionization,
star formation and stellar feedback. Interactions of a satellite
galaxy with its host are included in the form of stellar stripping and
tidal disruption as well as ram-pressure stripping, an effect that
leads to the removal of the hot gas reservoir from satellites
after infall.

As discussed by \citet{Starkenburg2013}, the simulated satellite
luminosity function of Aquarius haloes is consistent with that of the
Milky Way. Luminous satellites populate a minority of the subhalo
population, preferentially the high-mass end. Indeed, by number, most
subhaloes have low mass and, according to the model, remain completely
``dark'' throughout their lifetime.

\subsection{Satellite sample} 
\label{SecSatSample}

The semi-analytic model assigns a stellar mass (or luminosity) to each
subhalo at the present time. We classify them as: (i) ``classical''
satellites (i.e., those brighter than $M_V=-8$); (ii) ``ultra-faint''
satellites (fainter than $M_V=-8$); and (iii) ``dark'' subhaloes (i.e.,
those with no stars). We shall hereafter use the term ``luminous
subhaloes'' to refer to classical and ultra-faint satellites combined.

This classification makes reference to the Milky Way, where the
``classical'' satellite population is expected to be complete within
the boundaries of the Galactic halo with the exception perhaps of the
``zone of avoidance'' created by dust absorption in the Galactic
disc. ``Ultra-faint'' satellites, on the other hand, have only
recently been discovered in Sloan Digital Sky Survey (SDSS)
data. Their inventory is far from complete and their spatial
distribution highly biased to relatively small nearby volumes in the
region surveyed by SDSS \citep{Koposov2008}. Because of this, we shall
restrict much of the comparison of our models with data on classical
satellites.

\begin{figure}
  \begin{center}
      \includegraphics[angle=270,width=0.5\textwidth]{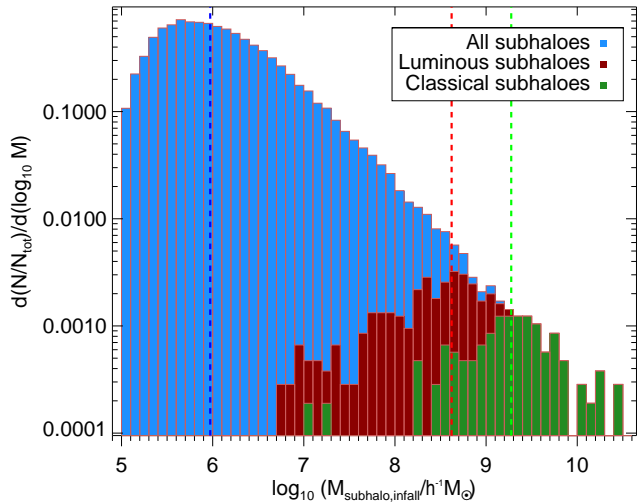}
  \end{center}
  \caption{Mass distribution of subhaloes found, at $z=0$, within the
    virial radius, $r_{200}$, of the level-2 Aquarius A through { F}
    haloes. Their (virial) masses are computed at the time of first infall
    into the main progenitor of the main halo. All subhaloes are shown
    in blue, luminous satellites in red, and classical satellites in green. Vertical dashed lines indicate the median of each
    group. Luminous satellites populate preferentially the high-mass
    end of the subhalo mass function. The decline in numbers below
    $\sim 10^6 \, h^{-1}\, M_\odot$ results from limited numerical
    resolution. We consider only subhaloes with masses exceeding $\sim
    10^6 \, h^{-1}\, M_\odot$ in our subsequent analysis.}
  \label{FigMsubDist}
\end{figure}

\section{Analysis and Results}
\label{SecResults}

\subsection{Satellite masses and radial distribution }

Fig.~\ref{FigMsubDist} shows the mass distribution of all subhaloes
identified at $z=0$ {\it within the virial radius}, $r_{200}$, of each
of the { six} Aquarius haloes considered here. Masses are quoted at the
time of first infall into the main progenitor of each halo ($t_{\rm
  inf}$), and correspond roughly to the maximum virial mass of each
subhalo prior to accretion. We also show in Fig. ~\ref{FigMsubDist}
the subhalo masses of the luminous satellites and confirm that, as
expected, they tend to populate the most massive subhaloes.

Low-mass subhaloes clearly dominate the numbers down to $10^6\,
M_\odot$, where the distribution peaks. The decline in numbers at
lower masses results from limited numerical resolution \citep[see][for
a detailed discussion]{Springel2008}. We shall therefore consider for
analysis only subhaloes with virial mass exceeding $10^6 \, M_\odot$ at
first infall, or haloes with more than $\sim 100$ particles.
Combining all { six} simulations, our full satellite sample consists of
 ${50,874}$ subhaloes, of which ${452}$ host luminous satellites: ${296}$
ultra-faint and ${156}$ classical dwarfs, respectively.

\begin{figure}
  \centering
      \includegraphics[angle=270,width=0.5\textwidth]{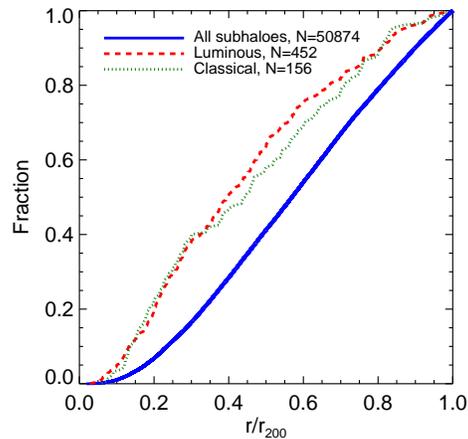}
      \caption{Fraction of enclosed subhaloes as a function of radius
        for level-2 Aquarius haloes A through { F}. All subhaloes are
        shown as a blue solid line; the subset of luminous satellites
        as a red dashed line, and only the classical as a green dotted
        line.}
  \label{FigCumRadDist}
\end{figure}

\begin{figure}
  \begin{center}
      \includegraphics[width=0.5\textwidth]{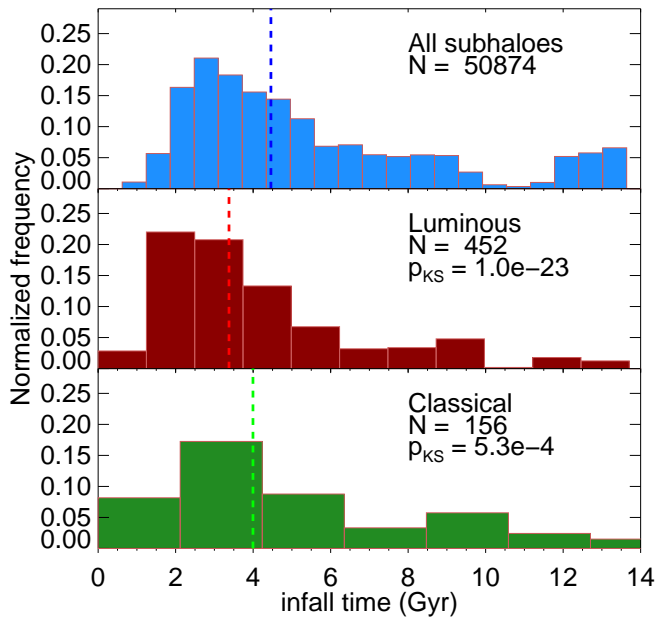}
  \end{center}
  \caption{Distribution of first-infall cosmic times { (where zero
    corresponds to the Big Bang)} for satellites identified within the
    virial radius of the main halo at $z=0$. Medians are indicated by
    vertical dashed lines. The normalization of the frequency is
    chosen such that the area under each histogram equals
    unity. Luminous (i.e., ultra-faint and classical) satellites enter
    the most massive progenitor of the main halo earlier than the
    average subhalo. $N$ indicates the number of subhaloes in each
    grouping. KS tests indicate the probability that the luminous or
    classical samples are drawn from the same parent population as all
    subhaloes.}
\label{FigHistTinf}
\end{figure}

Fig.~\ref{FigCumRadDist} shows the radial distribution of the three
populations of subhaloes in our model. Luminous satellites are
noticeably more centrally concentrated than the majority of subhaloes
\citep[e.g.,][]{Gao2004, Starkenburg2013}, a bias that might affect
the comparison between the orbital properties of luminous and dark
subhaloes.  Another noticeable difference between the luminous and
non-luminous subhalo population is the distribution of their infall
times, $t_{\rm inf}$. As shown in Fig.~\ref{FigHistTinf}, the luminous
subhaloes tend to fall in earlier.

\subsection{Orbital ellipticity distributions}
\label{SecEllipticity}

We compute the ellipticity, $e$, of the orbit of each subhalo from its
current apocentric, $r_a$, and pericentric, $r_p$, distances,
\be
e \equiv \frac{r_a - r_p}{r_a + r_p},
\ee
using the virial mass and concentration of the main halo. The
calculation assumes that the halo mass profile follows the NFW
\citep[NFW,][]{Navarro1996,Navarro1997} formula, where the
gravitational potential is written as
\be
\Phi(r) = -4 \pi G \rho_s r_s^2 \ \frac{\ln{(1+r/r_s)}}{r/r_s}.
\ee
Here $r$ is the distance from the centre of the main halo, and $r_s$
and $\rho_s$ are the NFW scale radius and density, respectively. The
scale radius, $r_s$, is related to the halo concentration by $r_s =
\r200 / c$, where $c$ is the NFW concentration parameter. The scale
density, on the other hand,  is related to the concentration parameter by
\be
{\rho_s \over \rho_{\rm crit}}= \frac{200}{3} \, \frac{c^3}{\ln(1+c)-c/(1+c)}.
\ee

\begin{figure}
  \begin{center}
      \includegraphics[width=0.5\textwidth]{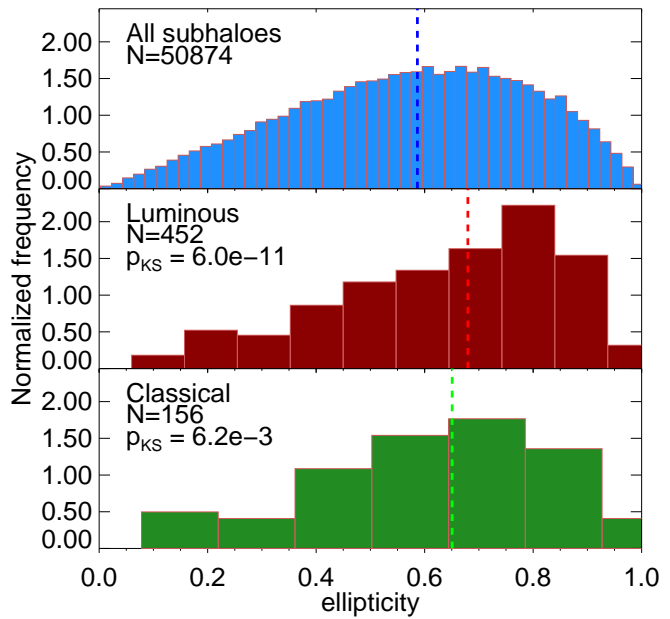}
  \end{center}
  \caption{As in Fig.~\ref{FigHistTinf}, but for the ellipticity
    distributions of all subhaloes (top); all luminous (middle); and just the
    classical dwarfs (bottom), measured at $z=0$. Medians are indicated by
    vertical dashed lines. KS tests indicate the probability that the
    observed luminous and classical satellite ellipticities are
    drawn from the same parent population as that of all subhaloes.}
  \label{FigEllDist}
\end{figure}

The ellipticity distributions of the three subhalo populations at
$z=0$ are shown in Fig.~\ref{FigEllDist}. The orbits of luminous
satellites are clearly more radial than those of the subhalo
population as a whole, which is dominated by the numerous low-mass,
``dark'' systems. Half of all subhaloes are on orbits with
${e<0.59}$, but the median $e$ is significantly larger for
luminous systems:  ${0.68}$ for all luminous and ${0.65}$ for classical
satellites.  As indicated by a Kolmogorov-Smirnov (KS) test, the
distributions are very significantly different indeed. (The
probability that the $e$-distribution of each satellite grouping is
drawn from the same parent distribution as all subhaloes is listed in
the middle and bottom panels.) This result is in qualitative agreement with pioneering work from \citet{Tormen1997} who found that within simulated cluster environments more massive satellites move along more eccentric orbits than lower mass satellites. 

Classical satellites are on slightly less radial orbits than
ultra-faints (as is reflected in the higher median $e$ for all
luminous subhaloes, compared to just the classical satellite subset),
but the difference between the two has lower significance; a KS test
yields a $p$-value of ${0.65}$.

We note that our modelling neglects the effect of baryons and, in
particular, of the potential modifications that the presence of a
massive stellar disc may have on the subhalo mass function and their
orbits. A recent study by \citet{Donghia2010}, for example, shows that
disc shocking may be able to destroy preferentially low-mass subhaloes
on plunging orbits. This would presumably skew their ellipticity
distribution to less radial orbits and would enhance the differences
noted above between the ellipticity
distributions of ``dark'' subhaloes and ``luminous'' satellites.

\subsection{Radial selection biases and dynamical evolution}
\label{SecSelBias}

What is the origin of the systematic differences in the orbital shapes
of luminous and dark subhaloes? 

A clue is provided by the distribution of infall times of all
subhaloes. As seen in the top panel of Fig.~\ref{FigHistTinf}, a
notable feature is that there is a well-defined dip in the number of
satellites with $t_{\rm inf}$ of the order of ${\sim 11}$ Gyr, followed by
a sharp upturn a couple of Gyr later. We have verified that the dip is
actually present in all Aquarius haloes taken individually, and does
not reflect a particular event in the accretion history of individual
haloes.

Rather, the dip may be traced to the fact that many subhaloes accreted
at $t_{\rm inf}\sim {11}$ Gyr are found temporarily outside the virial
boundary of the halo at $z=0$. Indeed, the radial period of an object
released at rest from the virial radius is roughly $\sim 3$ Gyr; most
systems accreted at $t_{\rm inf}\sim 11$ Gyr have apocentric radii
that exceed $r_{200}$, and the majority of them are today therefore
beyond the formal virial boundary of the halo. As discussed in detail
by \citet{Ludlow2009} \citep[see
also][]{Balogh2000,Mamon2004,Gill2005,Diemand2007,Ludlow2009,Wang2009},
subhaloes identified within the virial radius represent a rather
incomplete census of the substructure physically related to a halo:
many ``associated\footnote{We denote as ``associated'' all subhaloes
  that survive to $z=0$ and were, at any time during their evolution,
  within the (evolving) virial radius of the main halo. The number of
  associated subhaloes nearly doubles the number within the virial
  radius: we identify ${89,079}$ associated subhaloes in all
   { six} level-$2$ Aquarius haloes.}'' subhaloes are found outside the
formal virial radius of a halo at any given time. The effect is
mass-dependent: associated subhaloes outside $r_{200}$ tend to be
preferentially low mass.

\begin{figure}
  \begin{center}
      \includegraphics[width=0.5\textwidth]{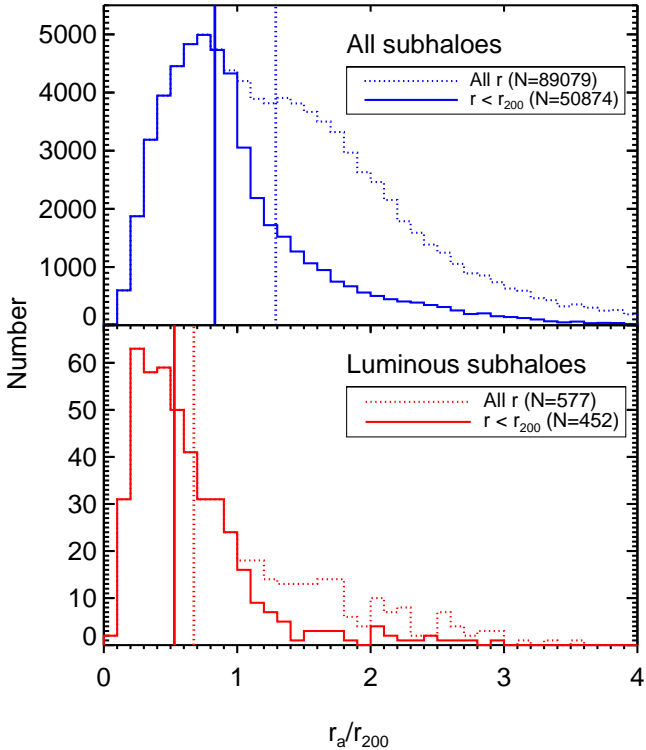}
  \end{center}
  \caption{Distribution of apocentric radii for all subhaloes and
    luminous subhaloes (top and bottom panels, respectively). Solid
    lines correspond to subhaloes found within $r_{200}$ at $z=0$;
    dotted lines to all ``associated'' subhaloes. Note that few ($\sim
    20\%$) luminous subhaloes are found outside the virial radius; on
    the other hand, selecting only systems within the virial radius
    excludes nearly half of all (mostly low-mass) associated
    subhaloes.
    \label{FigHistRapo}}
\end{figure}

\begin{figure}
 \begin{center}
      \includegraphics[width=0.5\textwidth]{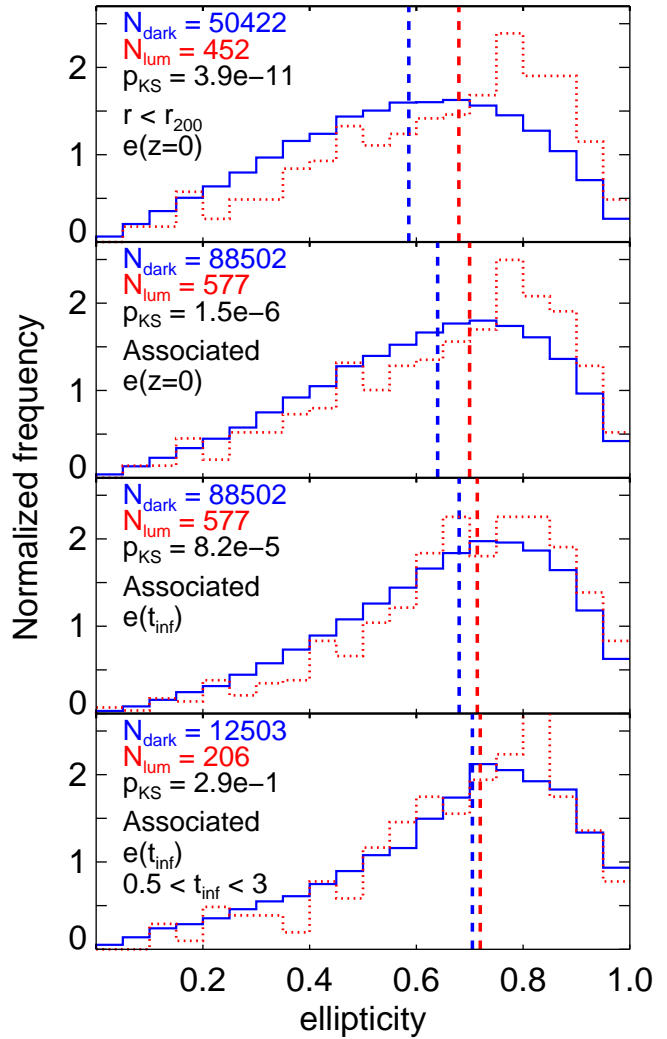}
  \end{center}
  \caption{Ellipticity distributions of luminous and non-luminous
    subhaloes (red dotted and blue solid lines respectively). Medians are indicated by
    vertical dashed lines. The listed $p$-values in each
panel correspond to the KS test between the luminous and non-luminous
subhalo distributions shown in that panel. The normalization is chosen such that the area under each curve is unity. {\it Top panel:} only subhaloes with $r<r_{200}$ are considered, their ellipticity distributions are measured at $z=0$. {\it Second panel:} Same as in the top panel, but including all ``associated'' subhaloes. {\it Third panel:} Same as in second panel, except that the ellipticity distribution is measured at the time of first infall. {\it Bottom panel:} Same as in third panel, but with the subhalo sample restricted to those subhaloes that have fallen into the main halo between $0.5$ and $3$ Gyr. \label{FigEllDistTinf}}
\end{figure}

We show this explicitly in Fig.~\ref{FigHistRapo}, where we compare
the apocentric radii of all associated subhaloes with those of
luminous ones. Selecting systems within $r_{200}$ includes more than
$80\%$ of all luminous associated satellites, but leaves out nearly
half of the less massive, dark subhaloes. This introduces a substantial
bias in the apocentric radii of the latter, selecting preferentially
systems with smaller apocentres. The effect on the orbital ellipticity
distribution is to favour systems with less radial orbits. 

This may be seen in Fig.~\ref{FigEllDistTinf}, where we compare the
ellipticity distributions of various samples of luminous and dark
subhaloes. The top two panels show that the dark and luminous subhalo
ellipticity distributions become more similar when considering all
associated subhaloes rather than selecting only those within
$r_{200}$. The radial selection bias, however, is not enough to
explain the systematic difference between the two populations, as
shown by the low probability of a KS test (see legends in each panel of
Fig.~\ref{FigEllDistTinf}). 

The remaining difference is { due partly to the fact that} the orbits of
dark and luminous subhaloes evolve differently after being accreted
into the main { halo. This is} shown in the third panel of
Fig.~\ref{FigEllDistTinf}, where ellipticities measured at the time of
first infall are compared. Although still significantly different,
ellipticities of associated dark and luminous subhaloes are
much closer at infall than at $z=0$.

The top three panels of Fig.~\ref{FigEllDistTinf} also indicate that
it is mainly the ellipticities of low-mass (dark) subhaloes that
change appreciably after infall: their orbits tend to become less
radial with time, something that is not seen in the luminous
satellites. Possible scenarios for this ``circularization'' of
low-mass subhaloes include the tidal dissolution of the groups to
which they belong at accretion, but also perturbations by massive
subhaloes they encounter on their orbits within the host halo
\citep[e.g.,][]{Tormen1998,Taffoni2003}.  { We have explicitly checked
that this conclusion is not the result of limited numerical
resolution: we obtain similar results even if we raise the minimum
subhalo mass considered in our sample from $10^6 M_\odot$ to $10^7
M_\odot$, or even $10^8 M_\odot$. }

Finally, { when comparing the ellipticities at infall, the difference
between dark and luminous subhaloes vanishes when considering systems
that were accreted in the same infall time window (bottom panel of
Fig.~\ref{FigEllDistTinf}). This is because satellites that fall in
early tend to be on slightly more radial orbits, as suggested by
\citet{Wetzel2011}. Selecting systems with similar infall times
removes this dependence and brings the ellipticity distribution of
dark and luminous subhaloes into agreement.}

We conclude that the orbital difference between dark and luminous
subhaloes shown in Fig.~\ref{FigEllDist} is due to the combined
effects of mass-dependent dynamical evolution after infall, a
{ dependence} of ellipticity with infall time, and by the selection bias
introduced by considering only systems within the virial radius.

\begin{table*}
\caption{Data for Milky Way satellites taken from the literature. Proper motions are given in equatorial coordinates; distances and velocities have been converted to a Galactocentric frame. References: 0=\citet{Piatek2005}, 1=\citet{Carrera2002}, 2=\citet{Walker2009b}, 3=\citet{Piatek2003}, 4=\citet{Pietrzynski2009}, 5=\citet{Walker2009a}, 6=\citet{Piatek2006}, 7=\citet{Pietrzynski2008}, 8=\citet{Piatek2007}, 9=\citet{Lepine2011}, 10=\citet{Bellazzini2005}, 11=\citet{Walker2007}, 12=\citet{Sohn2013}, 13=\citet{Bellazzini2004b}, 14=\citet{Mateo2008}, 15=\citet{Piatek2008}, 16=\citet{Udalski1999}, 17=\citet{Harris2006}, 18=\citet{Clementini2003}, 19=\citet{vanderMarel2002}, 20=\citet{Pryor2010}, 21=\citet{Monaco2004}, 22=\citet{Ibata1994} }
\centering  
\begin{tabular}{lccccccl} 
\hline\hline 

      &       &$\mu_\alpha$&$\mu_\delta$& $D_{MW}$ & $V_r$ & $V_t$ & \\ [-1ex] 
Galaxy& $M_V$ & (mas/century)&(mas/century)& (kpc) & (km/s) & (km/s) & References\\
\hline                  

Ursa minor&-8.8&-50.0 $\pm$ 17.0&22.0 $\pm$ 16.0&78 $\pm$ 3&-58.5 $\pm$ 6.4&157.8 $\pm$ 54.8&0,1,2\\
Carina&-9.1&22.0 $\pm$ 9.0&15.0 $\pm$ 9.0&107 $\pm$ 6&-4.8 $\pm$ 3.9&94.9 $\pm$ 40.1&3,4,5\\
Sculptor&-11.1&9.0 $\pm$ 13.0&2.0 $\pm$ 13.0&86 $\pm$ 6&78.0 $\pm$ 5.1&243.8 $\pm$ 52.9&6,7,5\\
Fornax&-13.4&47.6 $\pm$ 4.6&-36.0 $\pm$ 4.1&149 $\pm$ 12&-38.8 $\pm$ 1.9&185.9 $\pm$ 45.3&8,4,5\\
Leo II&-9.8&10.4 $\pm$ 11.3&-3.3 $\pm$ 11.5&236 $\pm$ 14&14.9 $\pm$ 4.3&312.4 $\pm$ 118.5&9,10,11\\
Leo I&-12.0&-11.4 $\pm$ 3.0&-12.6 $\pm$ 2.9&258 $\pm$ 15&167.6 $\pm$ 1.6&106.6 $\pm$ 34.0&12,13,14\\
SMC&-16.8&75.4 $\pm$ 6.1&-125.2 $\pm$ 5.8&61 $\pm$ 4&-9.8 $\pm$ 2.8&256.3 $\pm$ 32.7&15,16,17\\
LMC&-18.1&195.6 $\pm$ 3.6&43.5 $\pm$ 3.6&50 $\pm$ 2&67.2 $\pm$ 4.0&342.5 $\pm$ 20.9&15,18,19\\
Sagittarius dSph&-13.5&-275.0 $\pm$ 20.0&-165.0 $\pm$ 22.0&18 $\pm$ 2&140.9 $\pm$ 3.9&274.2 $\pm$ 32.7&20,21,22\\

\hline 
\end{tabular}
\label{TabPM} 
\end{table*}

\section{Application to the Milky Way}
\label{SecCompMW}

One main conclusion of the previous analysis is that cosmological
simulations make well-defined predictions for the ellipticity
distribution of satellite galaxies. These predictions can not be
compared directly to observations because the only available data are
instantaneous positions and velocities for those satellites with
distance, radial velocity, and proper motion estimates. A literature
search yields { such} data for nine of the thirteen Milky Way satellites
brighter than $M_V=-8$. These positions and velocities may be used to
estimate orbital ellipticities after assuming a mass profile for the
Galaxy. This allows us to place constraints on the total mass of the
Galaxy by requiring that the ellipticity distribution matches that of
simulated luminous satellites. We pursue this idea in
Sec.~\ref{SecMassMW}, after presenting the observational data set we
use in Sec.~\ref{SecSatData}.

\begin{figure}
  \begin{center}
      \includegraphics[width=0.5\textwidth]{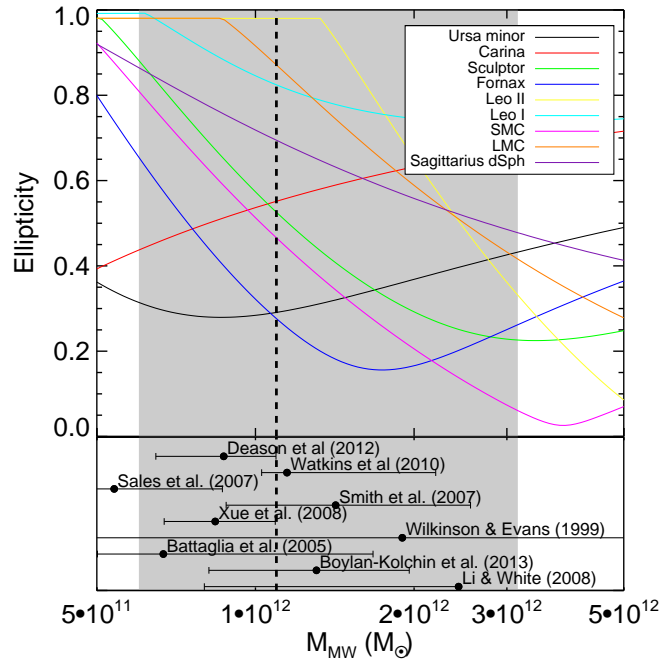}
  \end{center}
  \caption{The ellipticity of { orbits computed} for all MW satellites
    with proper motion measurements (given in Table~\ref{TabPM}) as a
    function of the assumed MW virial mass. The vertical dashed line
    gives the most likely MW mass as determined in this work and the
    grey area the 95\% confidence limits. { The bottom panel shows MW
    virial mass estimates from various literature sources, converted
    to $M_{200}$ (see text for details).} }
  \label{FigMassMW}
\end{figure}

\subsection{Milky Way satellite ellipticities}
\label{SecSatData}

We summarize in Table~\ref{TabPM} the Milky Way satellite literature
data used in our analysis. When several different estimates
are available we have adopted values from the recent compilation of
\citet{McConnachie2012}. We have exclusively adopted proper motion
estimates from HST data.

In order to facilitate comparison between observation and simulation
we have transformed all values to a Cartesian Galactocentric
coordinate system, with the $x$-axis pointing in the direction from
the Sun to the Galactic Centre, $y$-axis pointing in the direction of
the Sun's orbit, and $z$-axis pointing towards the Galactic North
Pole. We assume a velocity of $V_0=239 \pm 5$ \kms \ for the
clockwise circular velocity of the local standard of rest
\citep[LSR;][]{McMillan2011}; $R_0 = 8.29 \pm 0.16$ kpc for the
distance from the Sun to the Galactic Centre, as well as $(U,V,W) =
(11.10 \pm 1.23, 12.24 \pm 2.05, 7.25 \pm 0.62)$ \kms \ for the Sun's
peculiar velocity with respect to the LSR from \citet{Schonrich2010}.

We compute ellipticities for all nine satellites assuming that the
mass profile of the Galaxy may be approximated by an NFW halo with
concentration given by the mass-concentration relation of
\citet{Neto2007}. We have explicitly verified that the results we
quote are insensitive to the exact value of the concentration: for
example, varying $c$ between $8$ and $17$ for a halo of virial mass
${M_{200}=1.1 \times 10^{12} \, M_\odot}$ leads to an average change in
the ellipticity of ${0.05}$ over all nine satellites. This variation
is much smaller than the uncertainty implied by the relatively
poor accuracy of the proper motion estimates.

The coloured lines in Fig.~\ref{FigMassMW} show how the ellipticities
estimated for each MW satellite change as the assumed mass of the
Milky Way is varied from $M_{200}=5\times 10^{11}$ to $5\times 10^{12}
\, M_\odot$. As anticipated in Sec.~\ref{SecIntro}, the ellipticity of
a satellite depends sensitively on the mass of the Galaxy. For
example, if a satellite's radial velocity is much smaller than its
tangential velocity, then it must be either close to apocentre or
pericentre. If near pericentre, increasing the Galaxy mass will
decrease its apocentric radius and make the orbital ellipticity
decrease. If near apocentre, then the pericentre decreases as the mass
increases, resulting in a more elliptical orbit instead\footnote{These
  comments assume that the satellite remains bound as the Galaxy mass
  changes. Note that Leo I, Leo II, and the Large Magellanic Cloud would be unbound if
  the Milky Way virial mass was less than $6\times 10^{11}$, $1.5 \times 10^{12}$,
  and $8\times 10^{11} M_\odot$, respectively.}.
  
We note that a number of previous studies have suggested a possible
connection between episodes of star formation history in satellites
and pericentric passages during their orbits around the MW
\citep[e.g.,][]{Mayer2007,Pasetto2011,Nichols2012}. The strong
dependence of the timing of such episodes on the assumed mass of the
Milky Way provides an interesting constraint. It would be particularly
interesting, for example, to see if the {\it same} Milky Way mass
leads to synchronized pericentric passages and star formation episodes
for a number of satellites, an issue we plan to address in future
work.

\begin{figure}
  \begin{center}
      \includegraphics[width=0.5\textwidth]{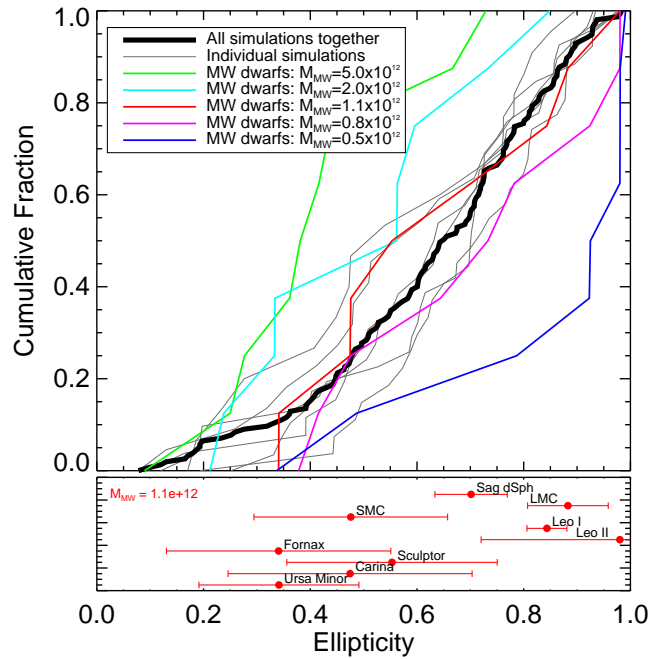}
  \end{center}
  \caption{{\it Top panel:} Cumulative ellipticity distributions of
    classical satellites from Aquarius haloes. Results for individual
    haloes are shown in thin grey; { that} for all six haloes combined
    is shown in thick black. Coloured lines show the ellipticity
    distribution of the nine classical dwarfs for which data are
    available, assuming different values for the virial mass of the
    Milky Way (see legends).  {\it Bottom panel:} Ellipticities
    estimated for each MW dwarf (arbitrarily offset in $y$ for
    clarity) assuming the best match halo virial mass for the Milky
    Way from this work, ${M_{200}=1.1\times 10^{12} M_\odot}$. One sigma error bars are
    given.  \label{FigCumEllip}}
\end{figure}

\subsection{The Mass of the Milky Way}
\label{SecMassMW}

As Fig.~\ref{FigMassMW} makes clear, in general the lower the total
mass the larger the inferred orbital ellipticity of a satellite. For example,
the median ellipticity of the nine MW satellites increases from $0.4$
to $0.98$ as the mass is varied in the range described above. The
corresponding cumulative ellipticity distribution for several distinct
choices of the Milky Way mass is shown in the top panel of
Fig.~\ref{FigCumEllip} (coloured lines) and compared with that of
{ classical} satellites in Aquarius (thin grey lines for individual haloes
and a thick black line for all haloes combined).

 Note that the ellipticity distributions of individual Aquarius
  haloes are very similar despite large differences in their accretion
  history and the fact that they span a sizeable mass range
  \citep{Springel2008}. Even Aq-F, which underwent a recent major
  merger and is thus an unlikely host for the Milky Way, is
  indistinguishable from the rest. We caution, however, that our
  analysis is based on only six haloes, which precludes a proper
  statistical study of the halo-to-halo scatter. Future simulations
  should be able to clarify this, as well as the possible dependence
  of satellite properties as a function of halo mass and
  environment. Encouragingly, our conclusion agrees with the earlier
  work by \citet{Gill2004}, who analyse eight simulations chosen to
  sample a variety of formation histories, ages and triaxialities and
  report a striking similarity in the ellipticity distribution of
  their satellite systems. Furthermore, \citet{Wetzel2011} find that
  the ellipticity distribution of satellites at $z=0$ is independent
  of host halo mass in systems less massive than $\sim 4 \times 10^{12}
  M_\odot$, a range that comfortably includes most current estimates
  of the MW virial mass.

We conclude that comparing MW satellite ellipticities with the
simulation predictions offers a viable alternative method for
estimating the Milky Way mass. The best match (as measured by the
maximum value of the KS probability obtained when comparing the nine
MW satellite ellipticities to all six Aquarius haloes) is obtained for
$M_{200}=1.1\times 10^{12} M_\odot$. Values less than ${6\times
  10^{11} \, M_\odot}$ or larger than ${3.1 \times 10^{12}\,
  M_\odot}$ are { disfavoured} at better than $95\%$ confidence
according to the same test.

The bottom panel of Fig.~\ref{FigCumEllip} shows the ellipticities for
all satellites for the favoured Milky Way mass including 1$\sigma$
error bars. \citet{Lux2010} show that measurements with Gaia's
expected accuracy will enable calculations of the last apo- and
pericentres of each orbit to an accuracy of $\sim$14\% for a given MW
potential, whereas current observational data only allow recovery to
$\sim40\%$ accuracy. The Gaia data set will thus greatly enhance the
accuracy of the Milky Way mass determination using this method.

In Fig.~\ref{FigMassMW} the currently favoured mass range is shown by
the black dashed line and grey shaded area. We compare in the same
figure our results with independent estimates based on a variety of
methods, from the timing argument \citep{Li2008}, to the kinematics of
halo stars \citep{Battaglia2005,Smith2007,Xue2008,Deason2012}, to
virial estimates based on satellite kinematics
\citep{Wilkinson1999,Battaglia2005,Sales2007a,Watkins2010,Boylan2013}.
{ When other mass definitions were used, the estimates given in these
papers have been converted to $M_{200}$ assuming NFW profiles with
concentrations computed from \citet{Neto2007}. Some of these values
require extrapolating masses measured within smaller radii out to the
virial radius. In spite of this, it is striking that all literature
values are} in reasonable agreement with our determination, lending
support to the viability of our method.

\subsection{The associated satellites of the Milky Way}

\begin{figure}
 \begin{center}
      \includegraphics[angle=270, width=0.5\textwidth]{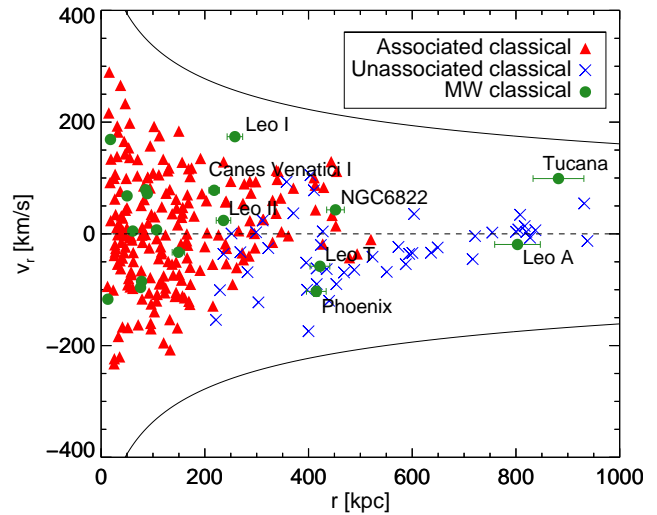}
  \end{center}
  \caption{Radial distance of classical subhaloes in the Aquarius simulations versus their
    radial velocity. All Aquarius subhalo data have been scaled to the best-fitting virial mass of $M_{200}=1.1\times 10^{12}
M_\odot$ as derived in the previous section. Red triangles correspond to the associated
    classical subhaloes (i.e. subhaloes that have at some time been inside the
    virial radius of the main halo) whereas blue crosses are
    classical subhaloes that have never been within the main halo. Overplotted
    in green filled circles are the classical dwarf galaxies within the Local Group that have the
    MW as their nearest massive neighbour (i.e. galaxies that
    are closer to M31 than to the MW are not included). { The solid
    lines show the escape velocity of the Milky Way, computed
    assuming an NFW halo with concentration equal to $8.52$. \label{FigVrVsR}}}
\end{figure}

As discussed in the previous section, a number of satellites
associated with the main halo are found today beyond the formal virial
boundaries of the halo. Although this applies mostly to low-mass
subhaloes, a small fraction ($\sim 20\%$) of luminous satellites
outside $r_{200}$ at $z=0$ have also been associated with the main
halo. Can we use their orbits to identify them? This is important
because associated satellites are more likely to have experienced
tidal and ram-pressure stripping and to have evolved differently from
``field'' dwarfs.

We explore this idea in Fig. \ref{FigVrVsR}, where we compare the
Galactocentric radial velocities and distances of ``classical'' dwarfs
in our model and in the vicinity of the Milky Way. The figure includes
only Local Group dwarfs that are closer to the MW than they are to the
Andromeda galaxy. All Aquarius main haloes have been normalized to
$M_{200}=1.1\times 10^{12} M_\odot$, our best match Milky Way mass as
determined in the previous section.  Associated model dwarfs are
plotted with red triangles and blue crosses denote dwarfs that have
never been associated with the main halo.

Interestingly, most subhaloes located between $\sim300$ kpc and
$\sim500$ kpc that are moving away from the main galaxy are
``associated'', whereas those with negative radial velocity tend to be
unassociated dwarfs on first infall. Furthermore, beyond $\sim600$ kpc
no classical dwarf has been associated with the main
halo. 

Some of these conclusions are in apparent conflict with the results
obtained by \citet{Teyssier2012}, who use associated subhaloes from
the Via Lactea II simulation { and report a significant population of
associated subhaloes out to $1.5$ Mpc from the host halo.} One
important reason for this apparent conflict is that
\citet{Teyssier2012} do not discriminate between subhaloes likely to
host a dwarf as bright as a ``classical'' dSph. Indeed, some
associated subhaloes { in Aquarius are also found beyond $1$ Mpc}, but
these are exclusively low-mass haloes unlikely, according to our
semi-analytical model, to host dwarfs brighter than $M_V=-8$.

We end with a word of caution, however. Our Aquarius main haloes do
not have a massive companion and the simulations therefore do not
attempt to reproduce the large scale distribution of matter of the
Local Group, where two massive haloes (those surrounding the MW and
M31) are about to collide for the first time. The main halo in the Via
Lactea II simulation does have a massive neighbour, which results in a
much larger turnaround radius for this system when compared to any of
the Aquarius systems. The effect of a Local Group environment on the
kinematics of outlying dwarfs has not been properly studied yet, but
there are indications that it is likely to play an important role in
the accretion history of satellite galaxies and in the evolution of
neighbouring dwarfs \citep[see, e.g.,][]{Benitez-Llambay2013}.

\section{Summary and Conclusions}
\label{SecConc}

We have combined the semi-analytical modelling of
\citet{Starkenburg2013} with the high-resolution simulations of the
Aquarius project to investigate the orbits of the satellites of
Milky-Way sized haloes in a $\Lambda$CDM universe. 

We find that the orbital ellipticity distribution of luminous
satellites shows little halo-to-halo scatter and is radially biased
relative to that of all subhaloes ``associated'' with the main
halo. The bias is relatively mild, considering that luminous
satellites populate preferentially massive subhaloes and are more
centrally concentrated than the main subhalo population. The bias
results from the combination of three main effects: (i) selecting subhalo
samples only within the virial radius; (ii) dynamical evolution
after infall; and (iii) a weak { dependence} of ellipticity with infall time.

The first arises because many low-mass subhaloes (which dominate by
number but are generally dark) have apocentric radii larger than the
virial radius and are thus found outside $r_{200}$ at any given
time. The second likely results from interactions between
substructures, which have a more pronounced effect on low-mass
subhaloes. Our results therefore urge caution when selecting only
subhaloes within the virial radius, since many associated subhaloes
(especially low-mass ones) lie at any given time outside the virial
radius. 

We have compared the ellipticity distribution predicted for luminous
satellites with that estimated for nine Milky Way satellites with
available 6D phase-space data. Since the latter depends sensitively on
the total mass assumed for the Milky Way, this comparison allows us to
place interesting constraints on the Milky Way mass. We find that the
ellipticity distribution of MW satellites is consistent with the
predicted one (at $95\%$ confidence) only if the MW virial mass is in
the range ${6\times 10^{11}}$-${3.1 \times 10^{12}\, M_\odot}$. This
determination is in agreement with current, independent constraints
from other observations, and is subject to improvement as the sample
of satellites with proper motion estimates increases and the accuracy
of such measurements improves.


\section*{Acknowledgments}
\label{sec:Acknow}
The authors are indebted to the Virgo Consortium, which was
responsible for designing and running the halo simulations of the
Aquarius Project. They are also grateful to Gabriella De Lucia, Amina
Helmi and Yang-Shyang Li for their role in developing the
semi-analytic model of galaxy formation used in this
paper. E.S. gratefully acknowledges the Canadian Institute for
Advanced Research (CIfAR) Global Scholar Academy and the Canadian
Institute for Theoretical Astrophysics (CITA) National Fellowship for
partial support.

\bibliographystyle{mn2e} 
\bibliography{paper} 

\label{lastpage}

\end{document}